\documentclass[aps,prd,reprint,secnumarabic,nonbibnotes,floatfix,amssymb,longbibliography,showpacs,preprintnumbers]{revtex4-1}

\usepackage{graphicx}
\usepackage{subfigure}
\usepackage{mathptmx}
\usepackage{helvet}
\usepackage{lineno}
\usepackage{textcomp}
\usepackage{multirow}
\usepackage{amsmath}

\begin{document}

\title{ Modeling of quantum effects in the hadronization }

\author{\v{S}\'{a}rka Todorova-Nov\'{a}}  
\affiliation{Institute of Particle and Nuclear  Physics, Charles University, Prague}
\email[sarka.todorova@cern.ch]{}

\begin{abstract}   
        
   A recent observation of a threshold momentum difference in the
   production of adjacent hadrons is implemented in the fragmentation model 
   of a three-dimensional QCD string, with the aim to investigate the
   common origin of the azimuthal ordering of hadrons and of the so-called Bose-Einstein correlations. 
   The role of particle decays and their polarization in the
   measurement of two-particle correlations is evaluated. A comparison with
   the available data is presented and further measurements
   suggested. The impact of the hadronization  on the long range angular
   correlations is discussed.
     
\end{abstract}

\pacs{13.66.Bc,13.85.Hd,13.87.Fh}

\maketitle


\section{Introduction}

   The model of the quantized fragmentation of the helical QCD string  ~\cite{hel2}  is built upon the observation
    that hadron mass spectra corresponds to the string breaking in quantized amount of transverse energy, under the
    causal constraint of a cross-talk between breaking vertices. Within the model, the masses of hadrons below
    1 GeV can be calculated from just two parameters describing the transverse shape of the QCD string
    ( the angular energy density $\kappa R$, where $\kappa\sim$ 1 GeV/fm is the linear energy density
    of the string and $R$ is the radius of the helix,  and the quantized helix phase difference $\Delta\Phi$ ).
    With the model parameters  constrained by the meson mass spectrum,  the model suggests there is a distinct
    correlation pattern between adjacent hadrons, with a threshold momentum difference. Using the
    symmetries stemming from the assumption of the local charge conservation in the string breaking, 
    such a behavior was observed by ATLAS Collaboration in the minimum bias data at the LHC ~\cite{chains}.
    The same analysis managed the isolate the source of Bose-Einstein (BE) correlations. The data suggest
    these correlations are carried by chains of adjacent ground-state pions. The experimental input
    provides an estimate of the production rate of such chains. 
    
      The aim of the current study is to explore the possibility to translate the experimental input into a MC description
    of these hadronization effects which are expected to be universal. Such a MC implementation would help
    to settle the questions stemming from various interpretations of the same data, and to validate or to disprove the model, eventually.

\section{Data driven modeling}

\begin{figure}[bh!]
\begin{center}
\includegraphics[width=0.45\textwidth]{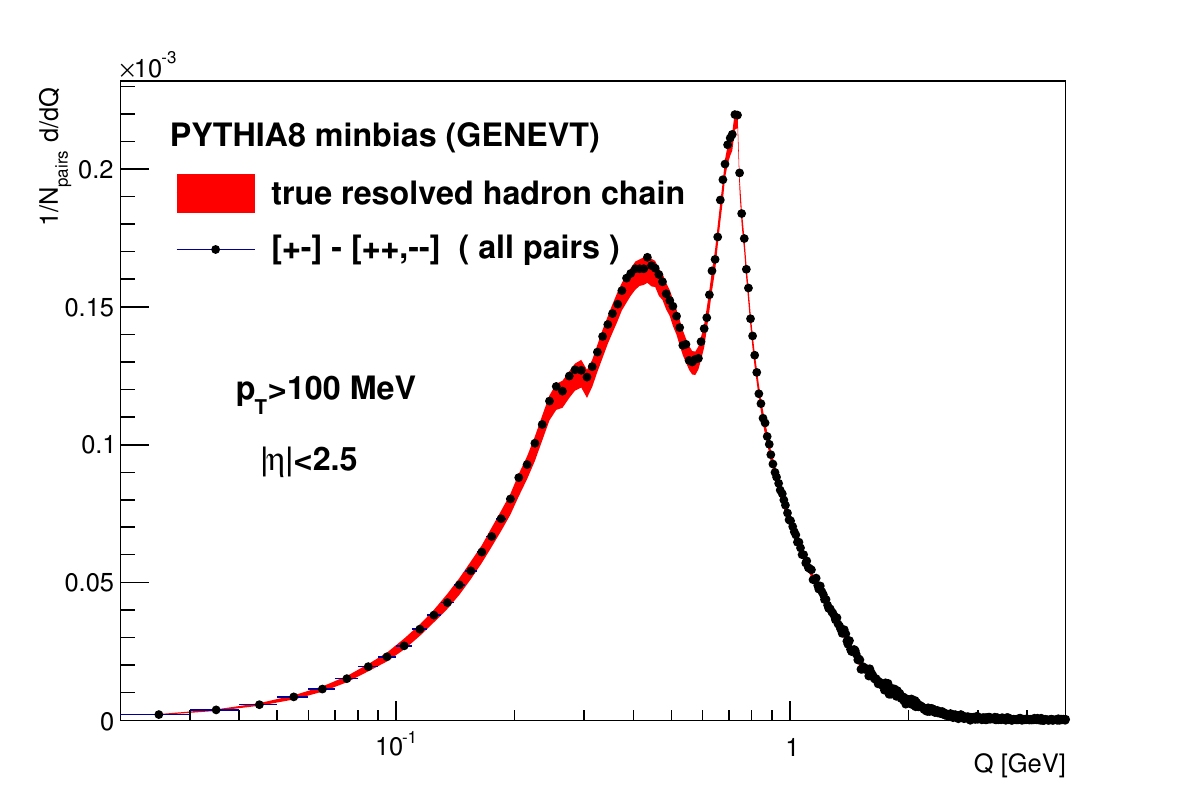}
\includegraphics[width=0.45\textwidth]{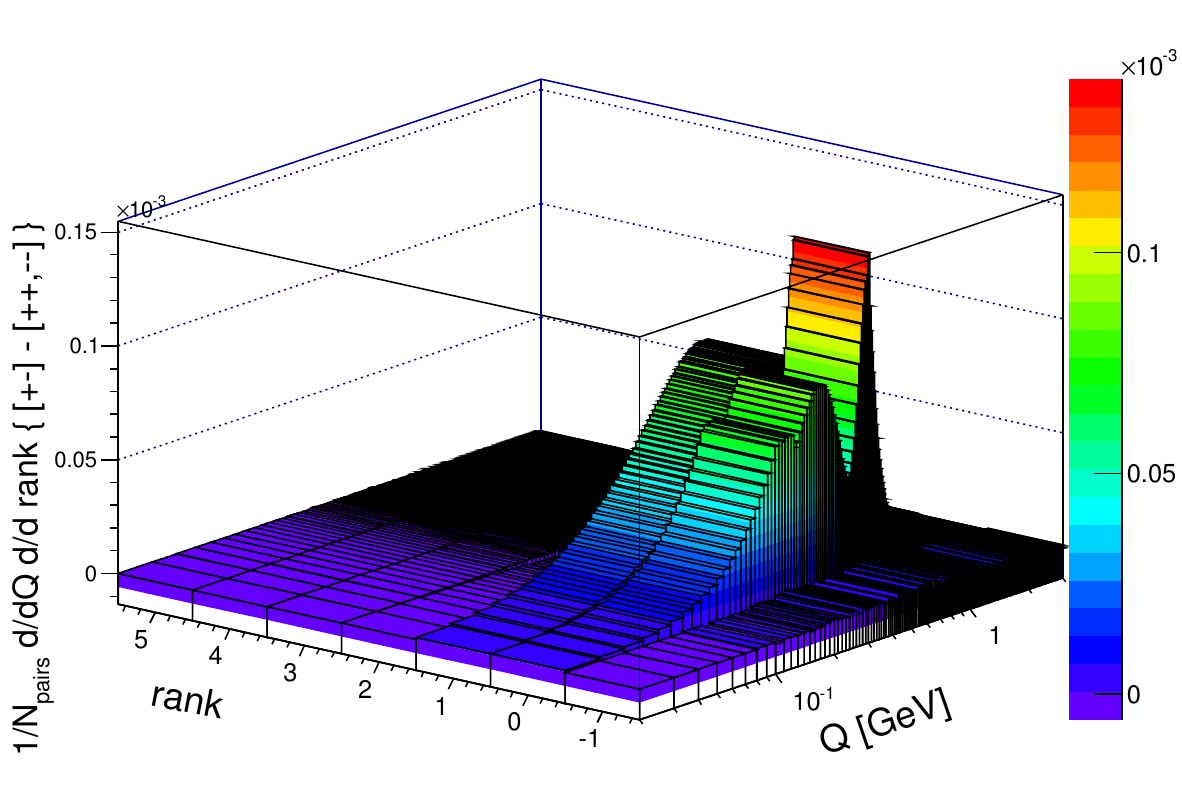}
\caption{ 
 Upper plot: Comparison of the Q spectrum obtained by the means of subtraction of like-sign pair distribution from that of opposite-sign pairs, with the
 true Q spectrum for adjacent charged hadrons, obtained from MC truth (the uncertainty marked by red band reflects the uncertainty in the resolving
 the ordering within a decay of a direct hadron).
 Bottom plot: Detailed view of the rank dependence of the subtraction between opposite-sign and like-sign pairs in MC; pairs of hadrons originating
 from different strings (color-disconnected chains) are assigned rank=-1. Rank 0 designs pairs of hadrons from the decay of a direct common resonance. 
\label{fig:subtraction_mc}}
\end{center}
\end{figure}

    The experimental technique which is particularly helpful in the study of hadronization effects uses the
    difference between pairs of particles with the opposite-sign charge and with the like-sign charge to measure the properties
    of adjacent pairs. The local-charge conservation forbids the creation of adjacent (rank 1) hadrons pairs with the same charge,
    and the random production of neutral hadrons makes the probability to produce a pair of hadrons with a specific
    charge combination about the same for rank difference 2 and more. It follows that the observable 
    
    \begin{equation}
       \Delta(Q)  =  \frac{1}{N_{\rm ch}}  \ [  \  N(Q)^{\rm OS}-N(Q)^{\rm LS} \ ]~,\\
    \end{equation}
       
        where $Q=\sqrt{-(p_i-p_j)^2}$ stands for the 4-momentum difference, should be particularly sensitive to the properties of pairs of adjacent hadrons. The number of adjacent hadron pairs rises linearly with the number of charged particles in the sample: the normalization of the $\Delta(Q)$ makes the integral
        of the distribution invariant ( $\int \Delta(Q)dQ$=0.5 in case of full acceptance measurement ).
        
         The assumptions made about $\Delta(Q)$ are verified in MC simulation where no correlations are included in the hadronization process
    beyond local charge and momentum conservation, and the pre-defined production rate of unstable hadrons with tabulated masses and widths.
    Fig.~\ref{fig:subtraction_mc} documents the subtraction of the combinatorial background for pairs with rank above 1 and for pairs formed by hadrons
    originating from different (color-disconnected) sources.  The non-direct hadrons inherits the rank of their ancestors in the notation used here.

\section{String fragmentation with helical string}

   It has been argued that in the production of pseudoscalar mesons ($\pi,\eta,\eta'$) and of chains of ground-state pions
 responsables for the enhanced production of close pairs of like-sign pions ($\pi^+\pi^-\pi^+,\pi^-\pi^+\pi^-$), the transverse
 mass and the longitudinal momentum decouple~\cite{hel2}. In the first approximation, the helical structure of the string
 can be implemented as a modulation on top of the default longitudinal string fragmentation (two degrees of freedom describing the generation of the intrinsic transverse momentum of hadrons are
 nevertheless suppressed as the transverse momentum of hadrons is obtained from the transverse shape of the string).
 Such a study can be performed by doing modifications in the Pythia generator~\cite{PYTHIA}. 
 In practice, all hadrons are assigned a quantized transverse mass (the nearest multiple of $\kappa R \Delta\Phi \simeq$ 0.192 GeV
 which is bigger than the tabulated mass). The size of the intrinsic
 transverse momentum being thus defined, its direction follows the rotation of the phase of the helical field. It is argued that it is the correlation introduced by the helical winding which is responsable
 for various observable correlation phenomena such as the azimuthal ordering of hadrons~\cite{atlas_ao} and the Bose-Einstein (BE) effect.

\section{Injection of ordered pion chains}

\begin{figure}[bh!]
\begin{center}
\includegraphics[width=0.45\textwidth]{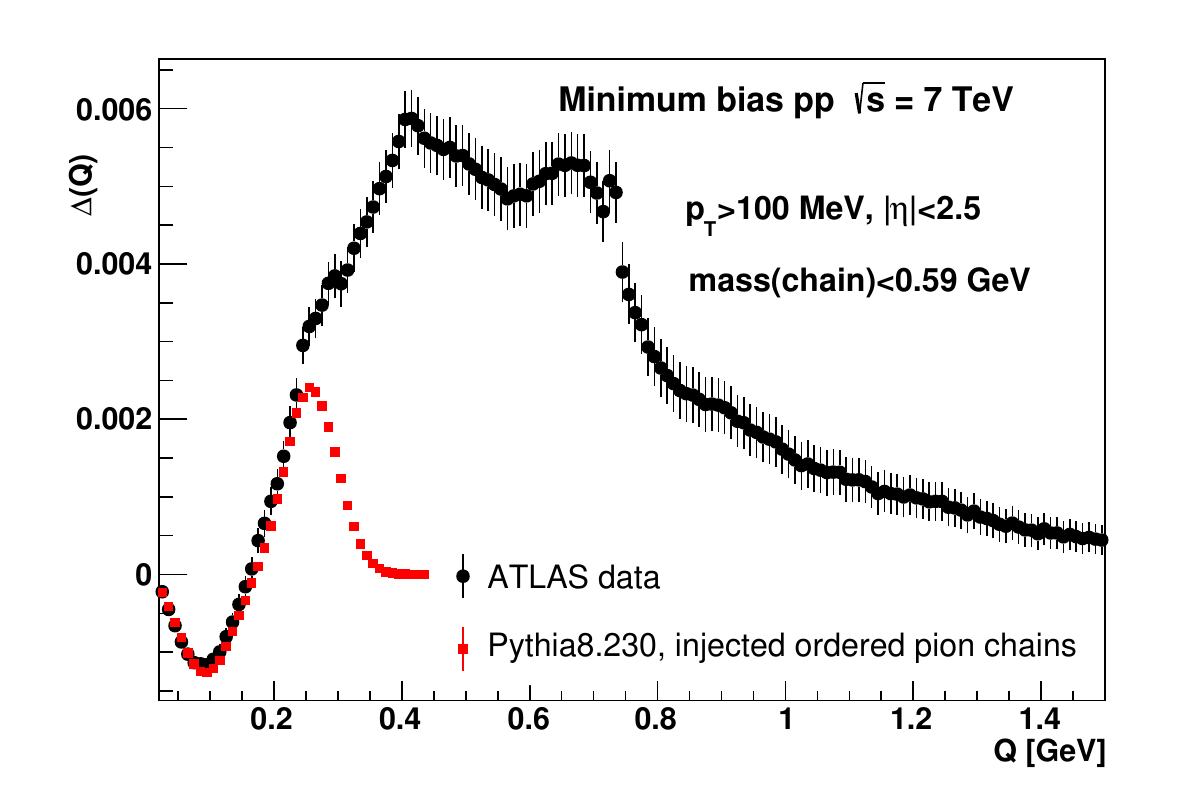}
\includegraphics[width=0.45\textwidth]{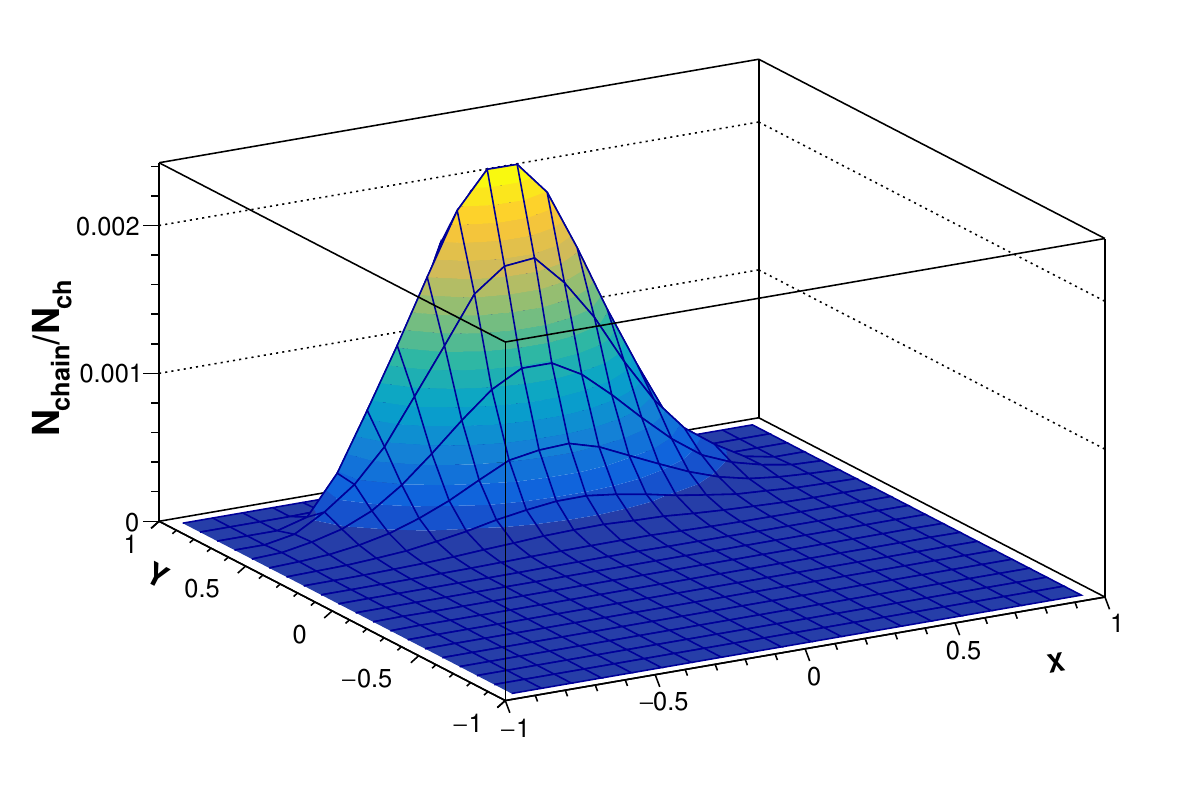}
\caption{ 
 Upper plot: The measured $\Delta(Q)$ ~\cite{chains}  and the shape generated by ordered charged pion triplet chains injected
  into simulation with rate of $\sim$ 1\% per charged hadron produced. 
   Bottom plot: Dalitz plot describing the 3-body decay pattern of injected triplets.
\label{fig:chains}}
\end{center}
\end{figure}

  The possibility that there is a well defined relation between the density of the helix winding and the energy of the emitting parton
 has not been explored yet ( such a relation would effectively replace the tuned parameters which are steering the longitudinal fragmentation
 function). Since the data suggest the BE correlations stems from part of string with approximately homogeneous helix field, or
 from a decay of heavier object (such as $\eta'$ meson), the ordered chains are injected into the fragmentation chains as triplets of pions with identical longitudinal momentum, correlated in the transverse plane. 
 The observed production rate of ordered triplets is about 1\% per final charged track when using
 the experimental acceptance region (p$_T>$100 MeV, $|\eta|<$2.5 ) as a reference. The correlations are encoded in the
 Dalitz parametrization of the three body decay of the triplet chain (the parametrization follows the one used in recent measurements of
 $\eta$ and $\omega$ decay~\cite{KLOE,WASA} ). Figure~\ref{fig:chains} shows the properties of the injected triplet pion chains: as in the
 measurement~\cite{chains}, the simulated chains describe the enhanced production of like-sign pairs at low $Q$.

\section{Hadron decays}

 \begin{figure}[h!]
\begin{center}
\includegraphics[width=0.45\textwidth]{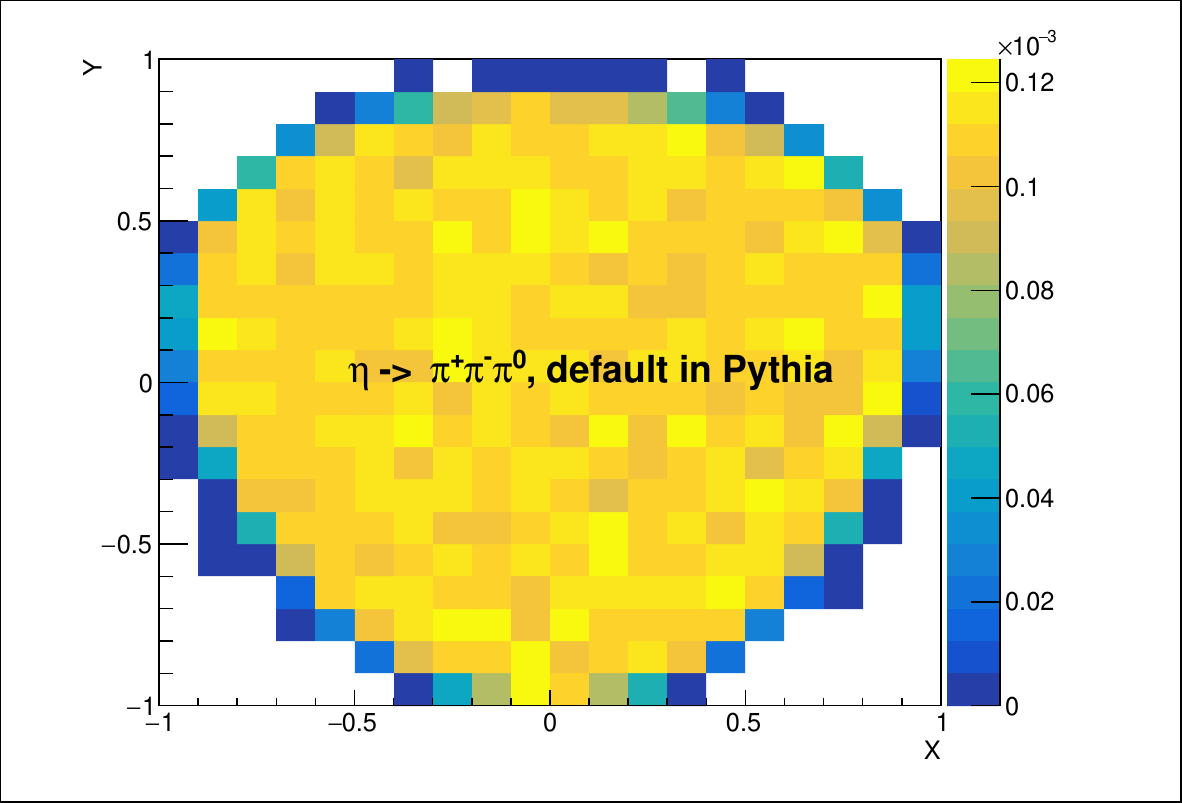}
\includegraphics[width=0.45\textwidth]{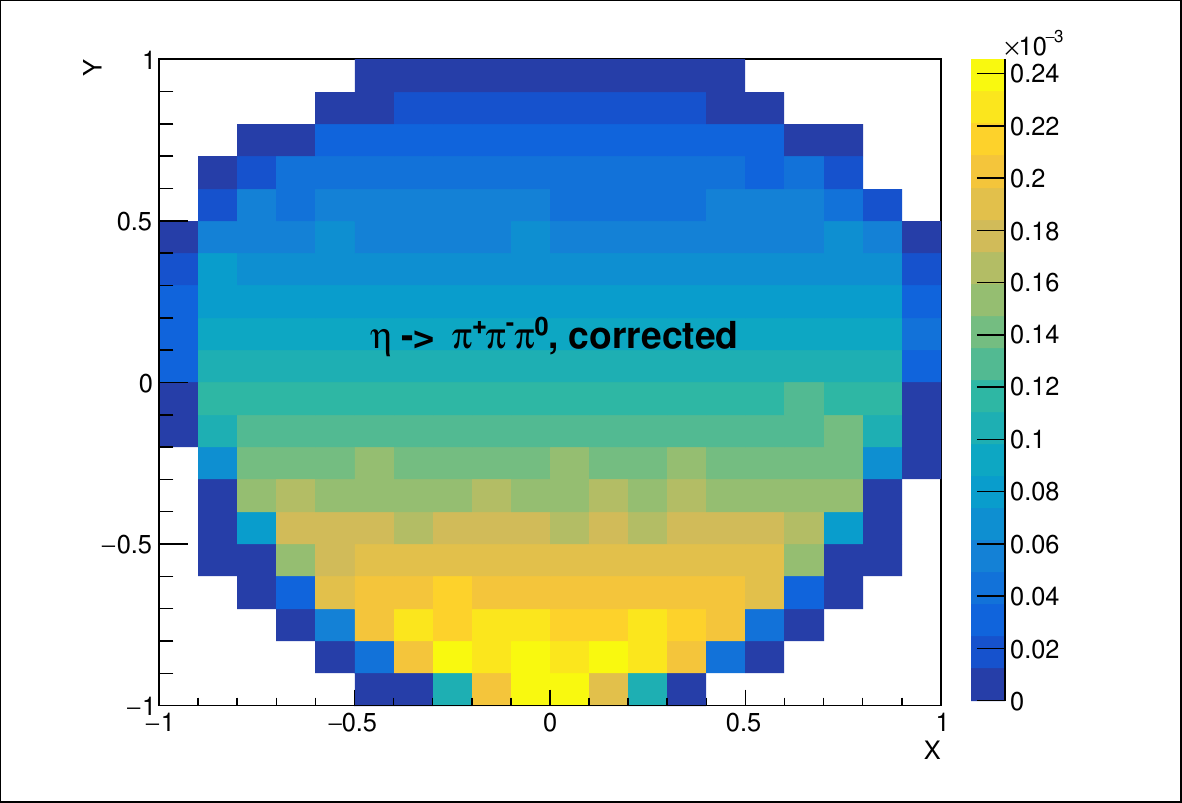}
\includegraphics[width=0.45\textwidth]{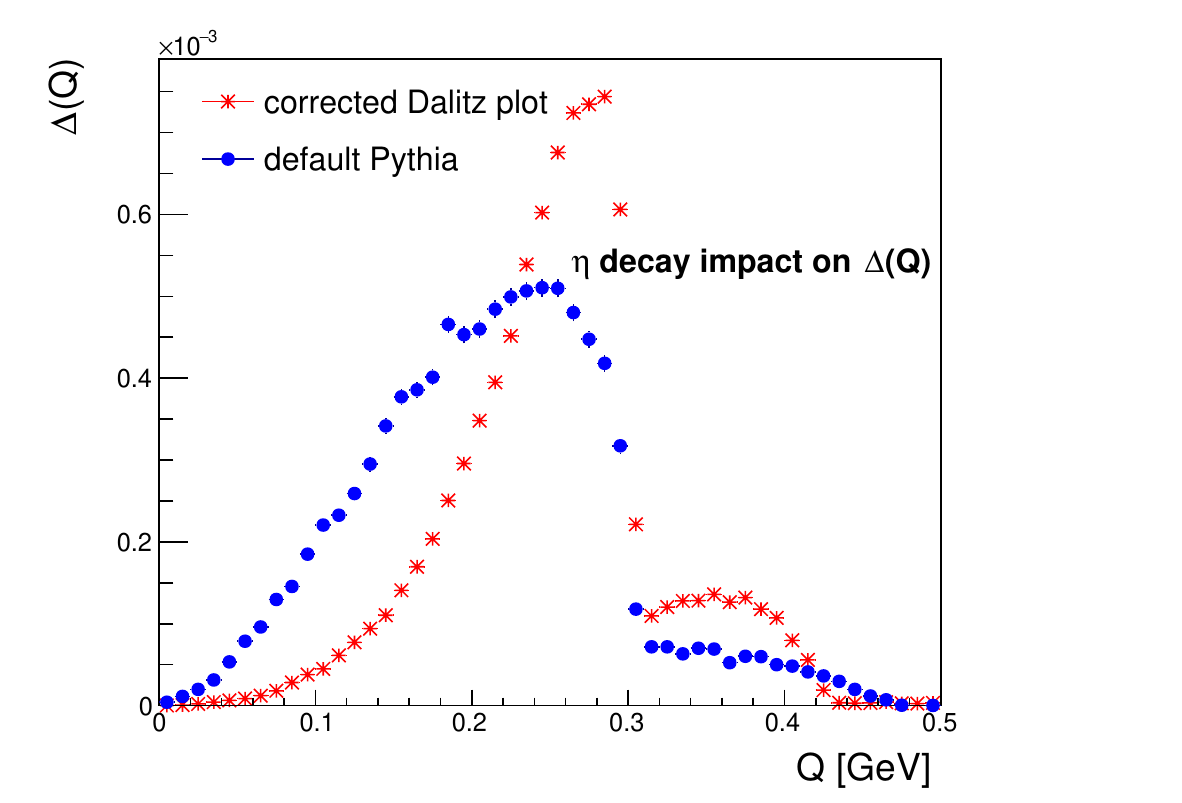}
\caption{ 
Top: The default 3-body decay pattern of $\eta$ in Pythia. 
Middle: The corrected 3-body decay pattern of $\eta$.
Bottom: The effect on the $\Delta(Q)$ distribution.
\label{fig:eta}}
\end{center}
\end{figure}
   
 \begin{figure}[h!]
\begin{center}
\includegraphics[width=0.45\textwidth]{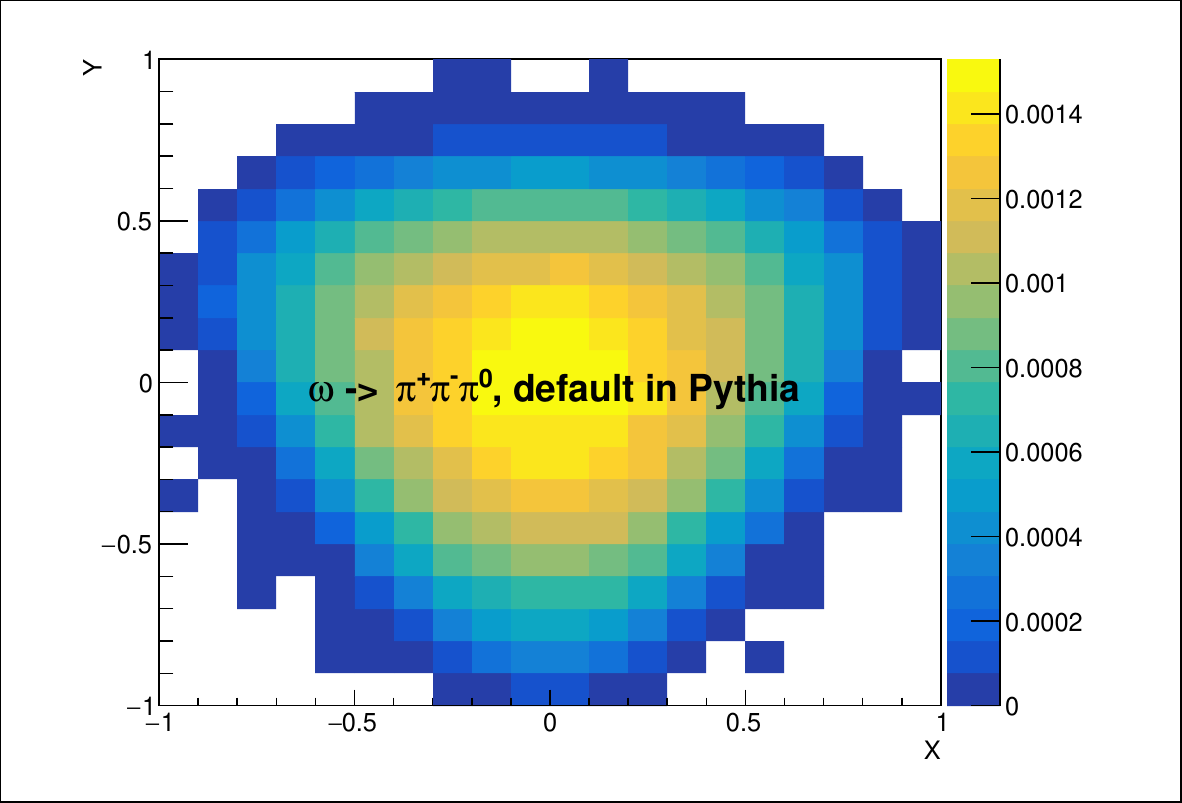}
\includegraphics[width=0.45\textwidth]{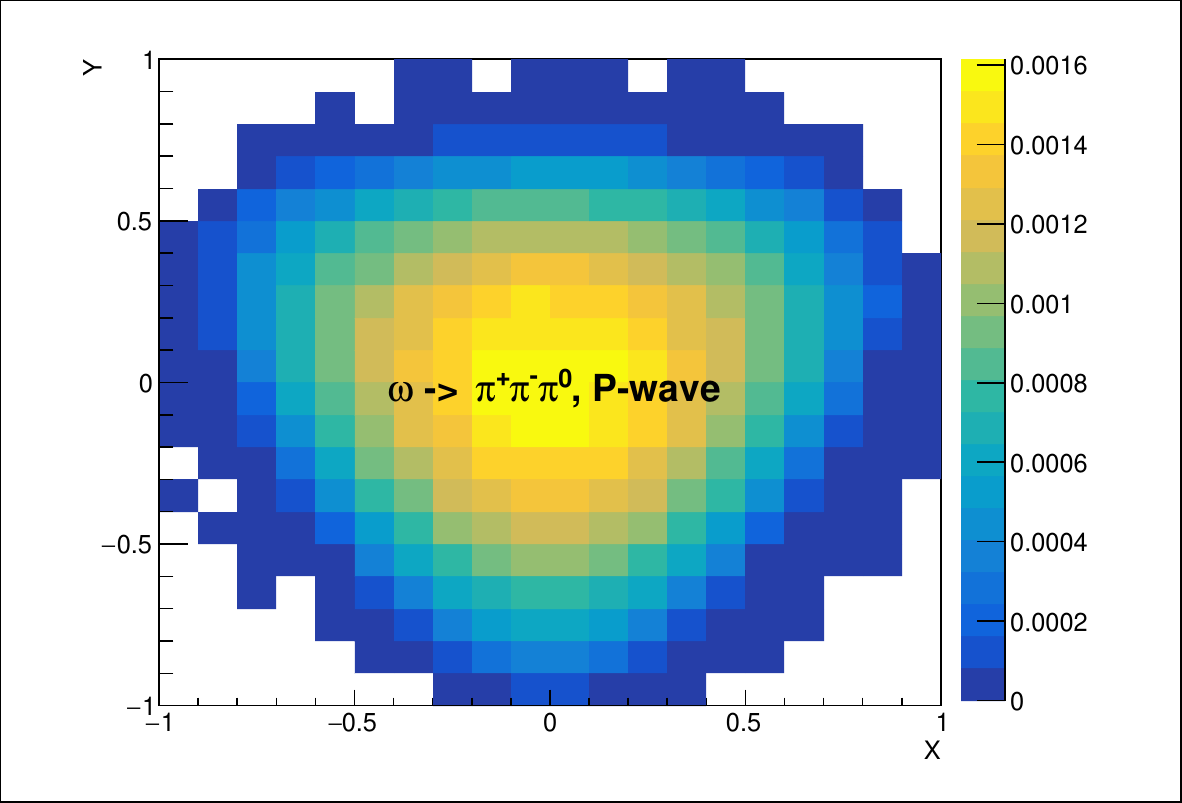}
\includegraphics[width=0.45\textwidth]{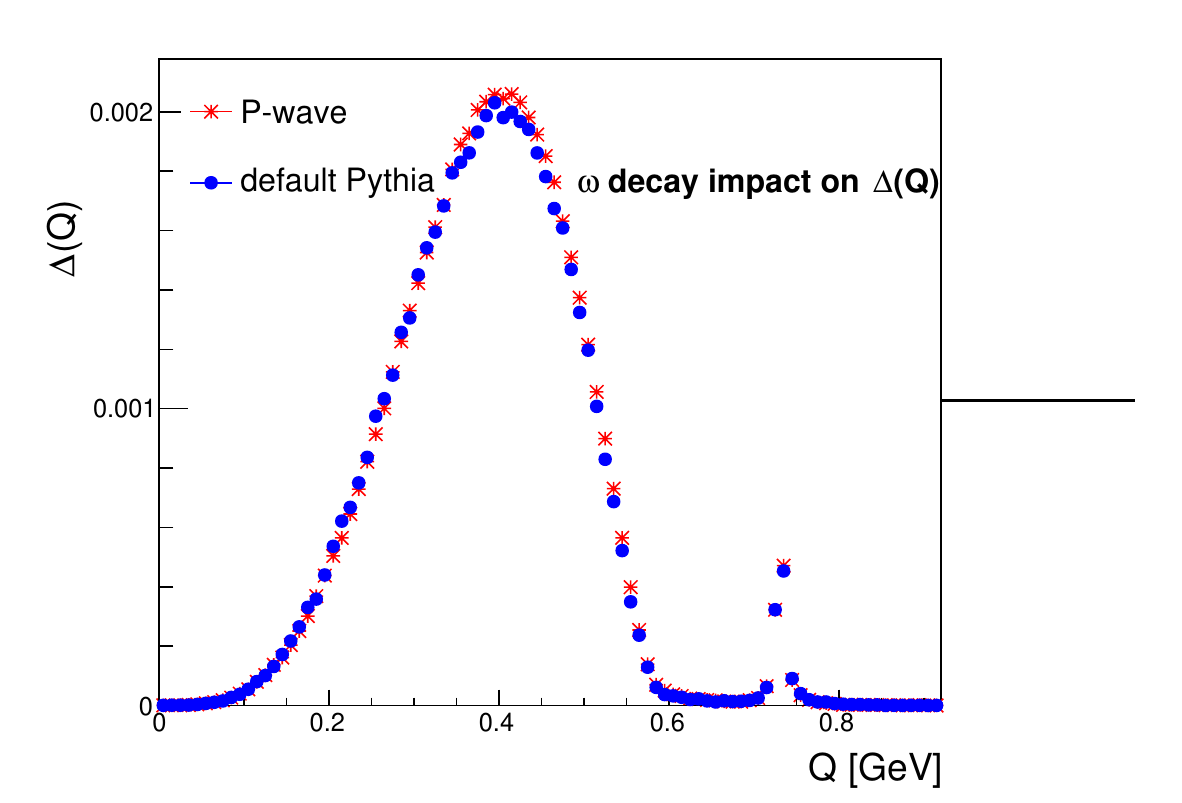}
\caption{ 
 Upper plot: The default 3-body decay pattern of $\omega$ in Pythia. 
 Middle plot: The P-wave distribution of $\omega$ decay.
Bottom: No modification is visible in the $\Delta(Q)$ distribution.
\label{fig:omega}}
\end{center}
\end{figure}

  For a proper description of the correlation pattern predicted by the model of helical QCD string, it is not sufficient to
 inject the source of correlations into simulation. The model predicts a threshold momentum difference for the production of adjacent hadrons
 at $Q\simeq$0.26 GeV  and effectively, looking at the Fig.~\ref{fig:chains}, there is no place left for additional adjacent hadrons
 in the region $Q<$0.25 GeV. In the standard Pythia simulation, this requirement is not fulfilled for pairs of hadron with rank 0 (particle decays)
 nor pairs of adjacent hadrons (rank 1, see Fig.~\ref{fig:subtraction_mc}). To ensure the predicted depletion of the low-Q region for adjacent hadrons
 is a task far more complex than the injection of ordered chains. Let's start the discussion of the problematics for the case of rank 0, i.e.
 the momentum difference between charged offspring of an unstable hadron.

   In the default Pythia simulation, the $\eta$ decay clearly violates the model prediction (Fig.~\ref{fig:eta}). However, it is known
 that the kinematics of the $\eta\rightarrow\pi^+\pi^-\pi^0$ decay does not correspond to the Pythia simulation ~\cite{KLOE}. After implementing
 the decay kinematics which agrees with the measured Dalitz plot occupation, the depletion of the low-Q region appears. A similar check
 of the $\omega$ decay using the P-wave dependence observed by ~\cite{WASA} shows that the default Pythia simulation is correct (Fig.~\ref{fig:omega}).
 It is not without interest to note that the resulting Q distributions of $\pi^+\pi^-$ pairs are very similar for $\eta$ and $\omega$, in the overlapping
 part of the spetrum.   

\begin{figure}[th!]
\begin{center}
\includegraphics[width=0.55\textwidth]{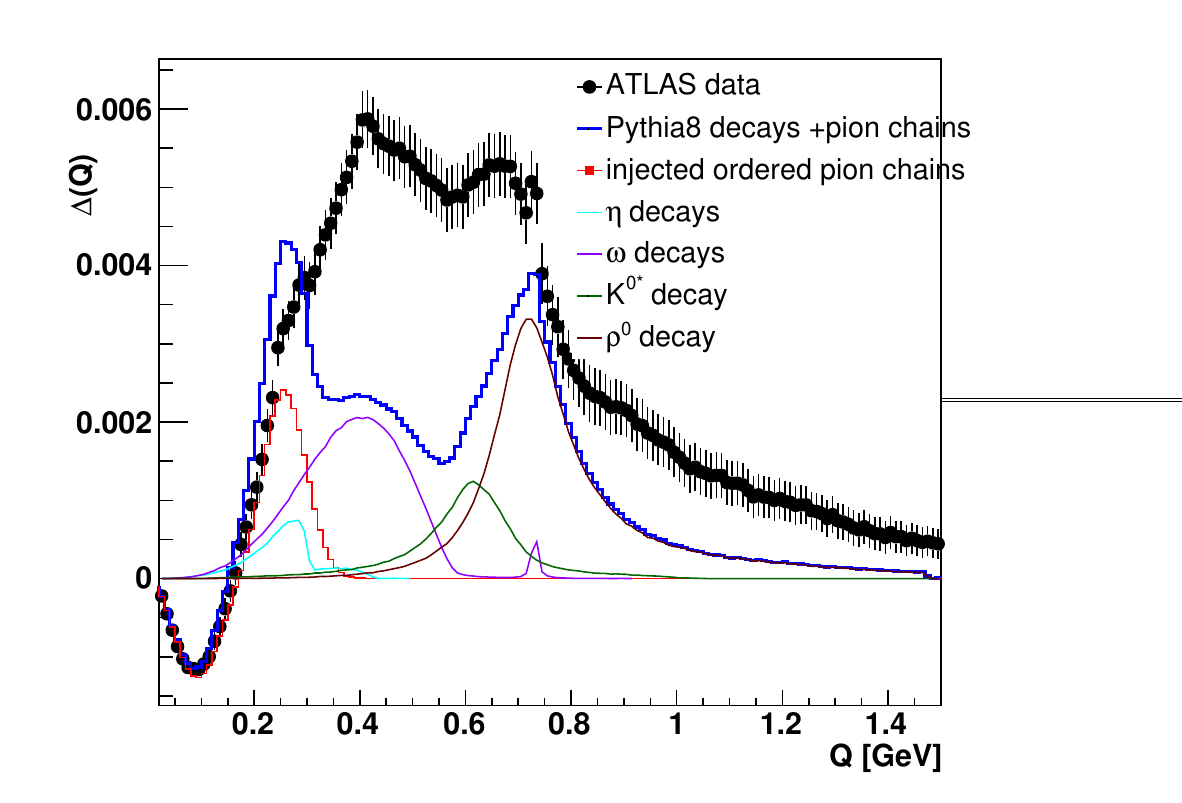}
\caption{ 
The contribution of pairs of hadrons with rank 0 to the shape of $\Delta(Q)$ in Pythia8.230.  A triplet chain is treated, technically, as an unstable direct hadron. 
\label{fig:direct}}
\end{center}
\end{figure}

 Fig.~\ref{fig:direct} shows the contribution from hadron decays
 (including the ordered chain decay and the $\eta$ correction) into the simulated $\Delta(Q)$ distribution.
 There is an excess in the simulation in the region of the
 $\eta$/chain decay. It is possible that the excess appears as the result of double counting of $\pi^+\pi^-$ pairs in the chain and $\eta$ decay contributions:
 for example, a decay of a $\eta'$ meson may contribute to both distributions.

\section{Adjacent hadrons, polarized decays }

\begin{figure}[bh!]
\begin{center}
\includegraphics[width=0.55\textwidth]{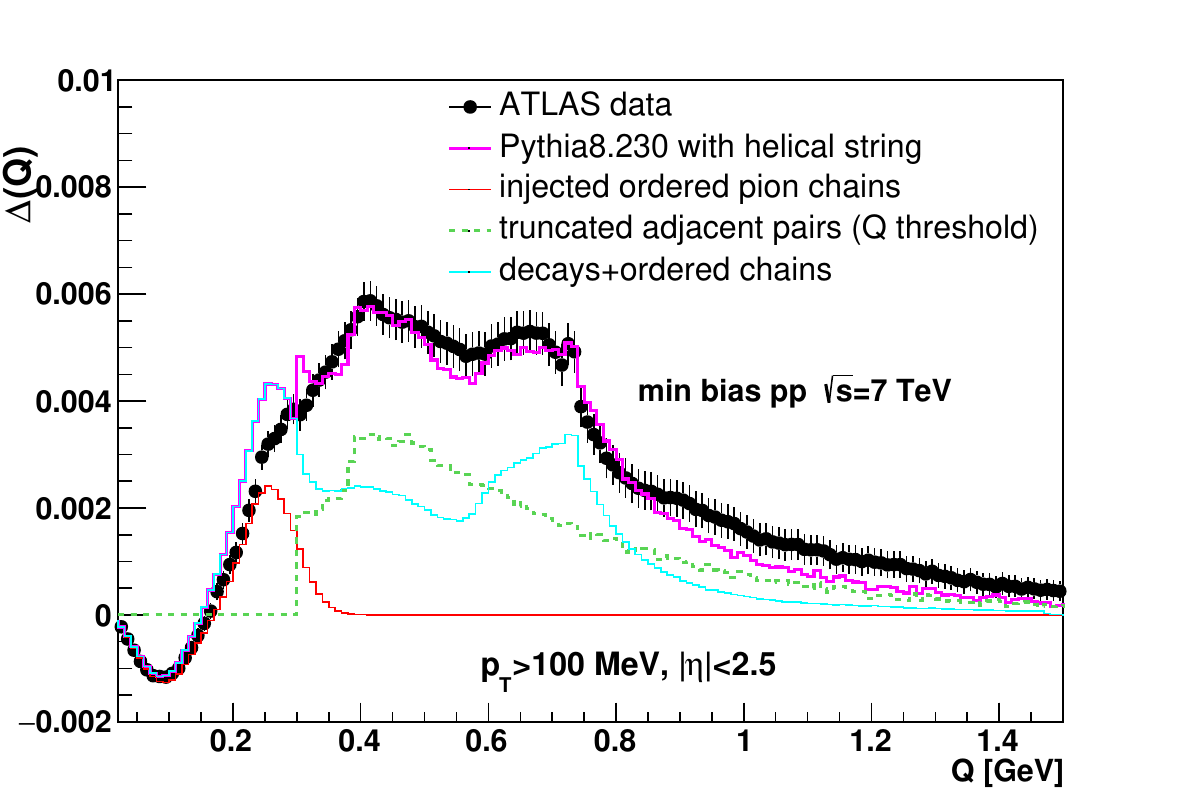}
\caption{ 
The contributions of direct and adjacent hadron pairs to the shape of $\Delta(Q)$ in Pythia8.230.  
The default $\rho^0$ production rate has been reduced by 20\%, the production rate of $K^{0*}$ has been increased by 50\%,
and $\rho^{+-}$ and $K^{+-*}$ decay plane has been rotated to the string transverse plane, in order to reproduce the overall shape of the data.
\label{fig:final}}
\end{center}
\end{figure}
 
  After a close inspection of combinations of particles of different origin, it seems the only possibility how to enforce the
 Q-threshold for adjacent hadrons, and to enhance the production of adjacent pairs in the region in between $\omega$ and $\rho$ peak ( both effects are correlated, since the integral of $\Delta(Q)$ is an invariant), is to properly simulate the polarization of decays of charged unstable hadrons.  
 A preliminary study of polarization effects in the decay of $\rho^{+,-}$ and $K^{*+.-}$ is encouraging, the rotation of their decay plane
 into the transverse region w.r.t. the string axis improves the agreement with the data. It is also obvious that a proper
 fool-proof algorithmic solution has yet to be developed, and further experimental input is necessary (see below).   
 In the meantime, the $Q$-threshold is enforced by an explicit veto on the remaining generated adjacent hadron pairs below $Q$= 0.3
 (shown in Fig.~\ref{fig:final}).

\section{Bose-Einstein correlations, azimuthal ordering - the same source?}

\begin{figure}[bh!]
\begin{center}
\includegraphics[width=0.5\textwidth]{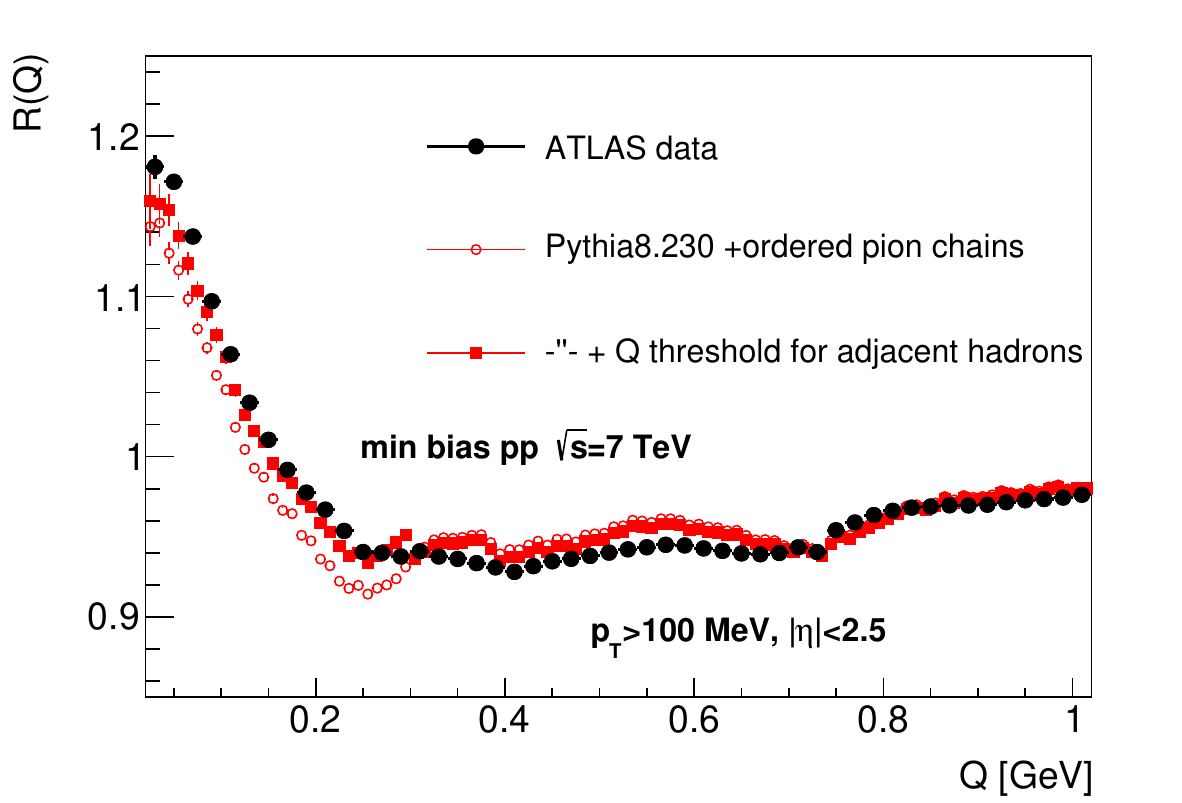}
\caption{ 
The injection of ordered pion chains translates into enhanced production of like-sign hadron pairs. $R(Q)$ is the ratio
 of the like-sign and opposite-sign inclusive Q distribution. The data are rather well reproduced, in particular
 after the enforcement of the Q-threshold for adjacent pair production.   
\label{fig:ratio}}
\end{center}
\end{figure}

   Figure~\ref{fig:ratio} shows the traditional correlation function, defined as the ratio of like-sign pair Q distribution
 and the unlike-sign pair Q distribution ( the latter is considered ``uncorrelated'' in the BE-motivated approach).
 The simulation based on the injection of correlated adjacent pion triplets describes the data well, in particular
 when the Q-threshold is enforced for all the adjacent hadrons. The role of the adjacent hadrons is primordial
 in the model and in the simulation, and it is not compatible with the picture of BE interference in the incoherent
 particle production the physicists usually work with. The assumptions made about the nature of the correlations
 have a large impact on the focus and interpretation of the experimental data : from the point of view of the
 hadronization studies, the ratio does not allow to extract the quantized properties of the hadron production
 (the gaussian shape
 of the excess of correlated like-sign pairs is completely invisible in the ratio of inclusive Q distributions).
 The ratio also retains a dependence on the shape of the uncorrelated combinatorial background, contrary to the
 subtraction, which removes the background efficiently. 

\begin{figure}[bh!]
\begin{center}
\includegraphics[width=0.5\textwidth]{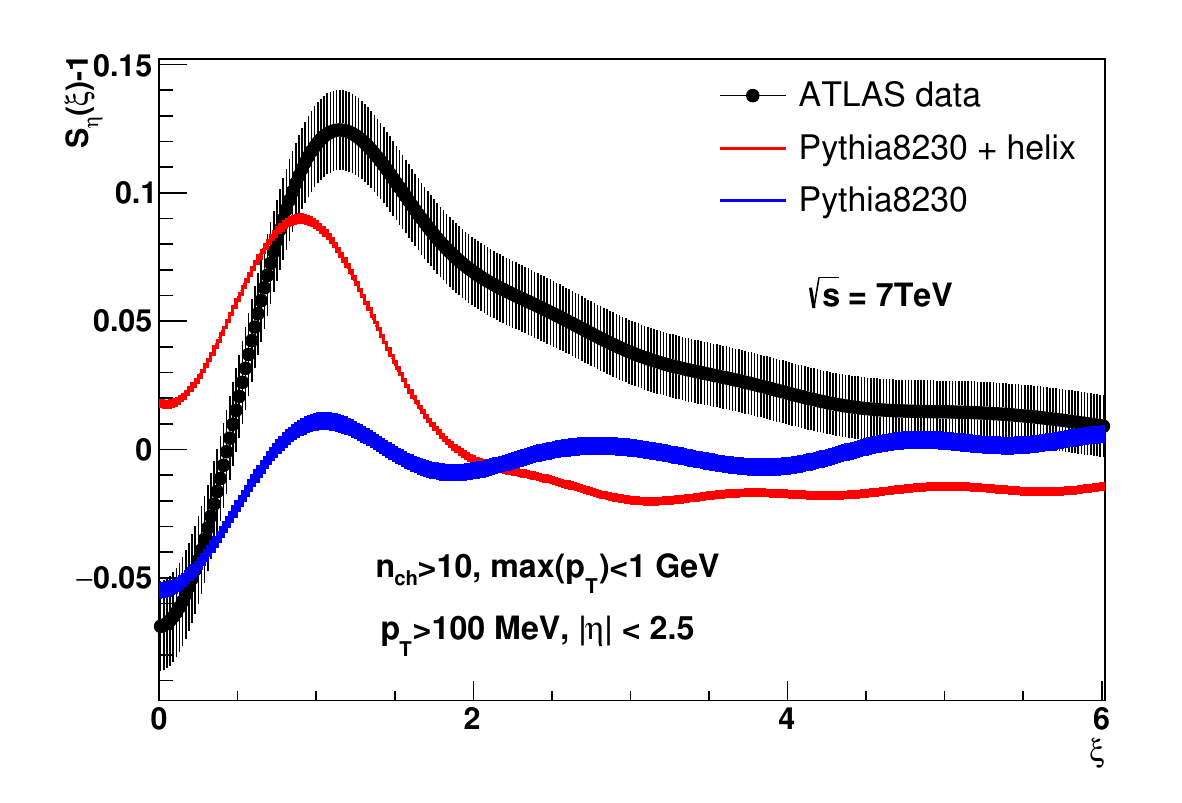}
\caption{ 
The injection of ordered pion chains translates into an azimuthal ordering signal.  The simulation is not corrected for the pairs of adjacent hadrons in the "forbidden" region below Q-threshold. The data are taken from~\cite{atlas_ao}. 
\label{fig:screw}}
\end{center}
\end{figure}

   Figure~\ref{fig:screw} shows the appearance of the azimuthal ordering (positive correlations in the power spectrum
measuring correlations between the azimuthal opening angle and the pseudorapidity difference is observed).  The data are not
 as well reproduced as the Bose-Einstein effect, but the hypothesis of the common origin
of both effects seems plausible. The measurement of the azimuthal ordering therefore provides a useful complementary
 input which should help to to adjust various components of the simulation.
 
 \begin{figure}[bh!]
\begin{center}
\includegraphics[width=0.5\textwidth]{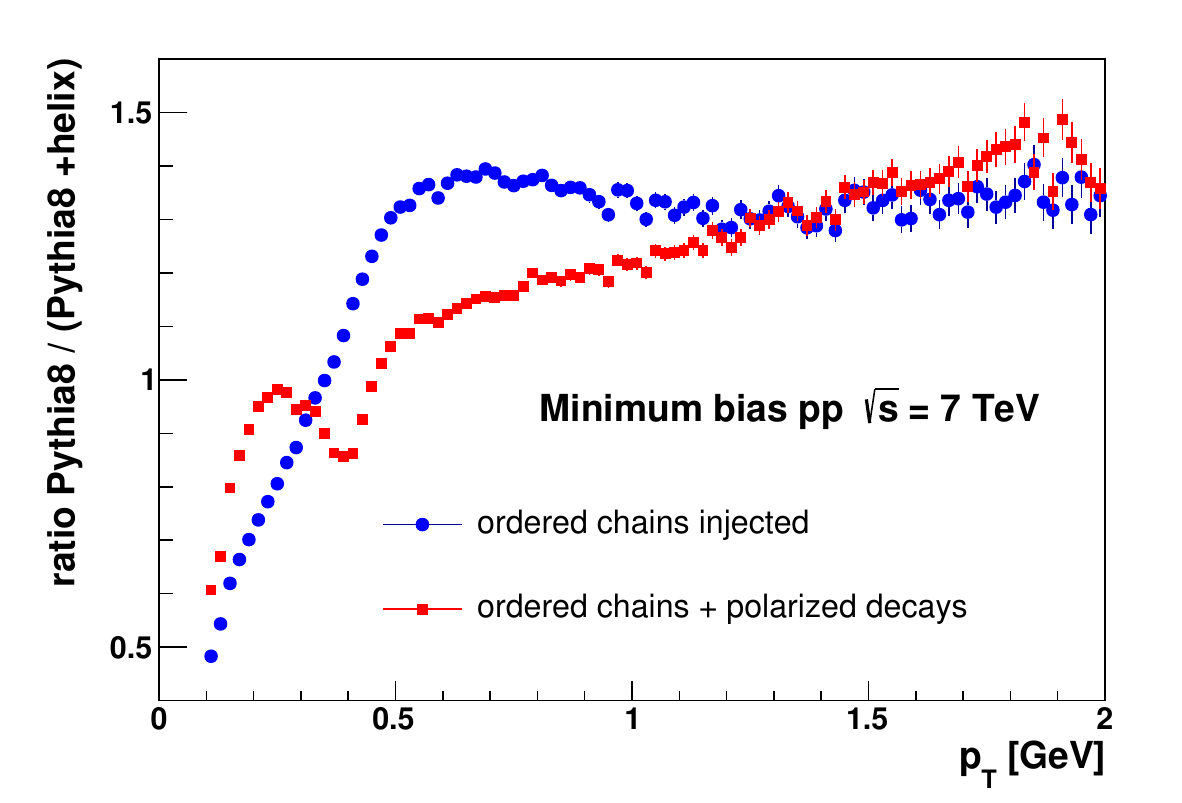}
\caption{ 
  The ratio of the default Pythia simulated inclusive $p_{T}$ spectrum, and the spectrum modified by helix string scenario,  for
 the minimum bias sample at $\sqrt{s}=$ 7 TeV. Blue points indicate the effect of the transverse momentum being derived
 from the transverse properties of the helical QCD string, red squares contain also the effect of the logitudinally
 polarized decay of $\rho$ and $K^*$ mesons.
\label{fig:pt}}
\end{center}
\end{figure}

\section{Inclusive transverse momentum spectra}

   The modifications brought by the helix string scenario have a significant impact on the shape of the
 inclusive transverse momentum distribution. Fig.~\ref{fig:pt} compares the default Pythia simulation
 with the enhacement brought by the narrow helix string scenario ( the measured $\kappa R\simeq$ 0.07 GeV)
 and with the effect mitigated by the polarization effects in $\rho$ and $K^*$ decay.

\section{Method of extraction of string parameters from measurement of correlation between adjacent hadrons}

 The model of quantized fragmentation of helical QCD string has the interesting property of being over-constrained, and there
 are multiple ways of measuring the basic string parameters, $\kappa R$ and $\Delta\Phi$.  They can be obtained from
 mass spectrum of light hadrons, but also, independently, from the measurement of correlations  between direct hadrons
 ordered according to the color flow history of the string fragmentation.  It is not possible to reconstruct the color flow
 from momenta of final hadrons only, as an exclusive measurement, but with help of just few well established symmetries 
 of the hadronization process, it is possible to do the measurement implicitly, for the specific case of chains of direct pions
 which represent the lightest color ordered hadron systems.
 
   In the pilot measurement of ordered hadron chains published by ATLAS ~\cite{chains}, which arrived at the result 
 which agrees with the prediction  based on hadron mass spectra, the longitudinal non-homogeneity of the QCD string has been neglected.
 
 The method which is outlined in following paragraphs is meant to overcome this limitation and to contribute to a more stringent
test of model predictions. 

  As a first step, the calculation of the momentum difference between adjacent hadrons is generalized in a way which allows
 to incorporate difference in the longitudinal component (along string axis) :
\begin{eqnarray}
   Q^2(p_a,p_b) & = & (\vec{p_{T_a} }- \vec{p_{T_b}})^2 + m^2_{T,a}  (\frac{z^+_b}{z^+_a}-1)   \nonumber  \\   
                             &  &   + m^2_{T,b}  (\frac{z^+_a}{z^+_b}-1),  
\end{eqnarray}
where $z_{a,b}$ stand for fraction of momenta of (one of two) generating partons taken by hadrons $a,b$.
 We can substitute, with some
approximation, the fraction $\frac{z^+_b}{z^+_a}$ with the experimentally accessible quantity $\zeta = min( \frac{|p_a|}{|p_b|}, \frac{|p_b|}{|p_a|})$ :
\begin{equation} \label{eq:plf_running}
  Q^2 \sim  (\vec{p_{T_a} }- \vec{p_{T_b}})^2 + m^2_{T}   ( \zeta(p_a,p_b) + 1/\zeta(p_a,p_b) - 2 ) , 
\end{equation} 
  where the quantization rule was applied: $m_{T,a} = m_{T,b} = m_{T} = \kappa R \Delta\Phi $ ( for pions ).  
  The approximation does not always work, since the momentum fraction is not, generally speaking,
  a boost invariant quantity, but nevertheless it is a good approximation for pairs of adjacent hadrons with common parton ancestors.
  
  A toy MC implementation (a slight modification of Pythia 8 code) is deployed where the hardest gluons in the parton shower are
  split to allow a simplified  $q\bar{q}$-like fragmentation to proceed without major impact on event shapes. The string pieces undergo a first
  round of standard Pythia fragmentation which defines the hadron composition. The resulting set of hadrons, ordered in color flow, is then reprocessed
  by quantized fragmentation which recalculates hadron momenta in the framework of helical QCD string.
  The exponential decay law similar to standard Lund technique is applied in the longitudinal string direction, with $b$ parameter
  randomly drawn from the interval (0.5-1).  As the procedure is solely intended to demonstrate the feasibility of such a measurement,
  no attempt is done to fine-tune the longitudinal properties of helical string decay ( no polarization effects introduced ).
  Naturally occurring chains of three direct charged pions are localized in the event output and analyzed as follows.

  \begin{figure} [bth]
  \begin{center}
  \includegraphics[width=0.45\textwidth]{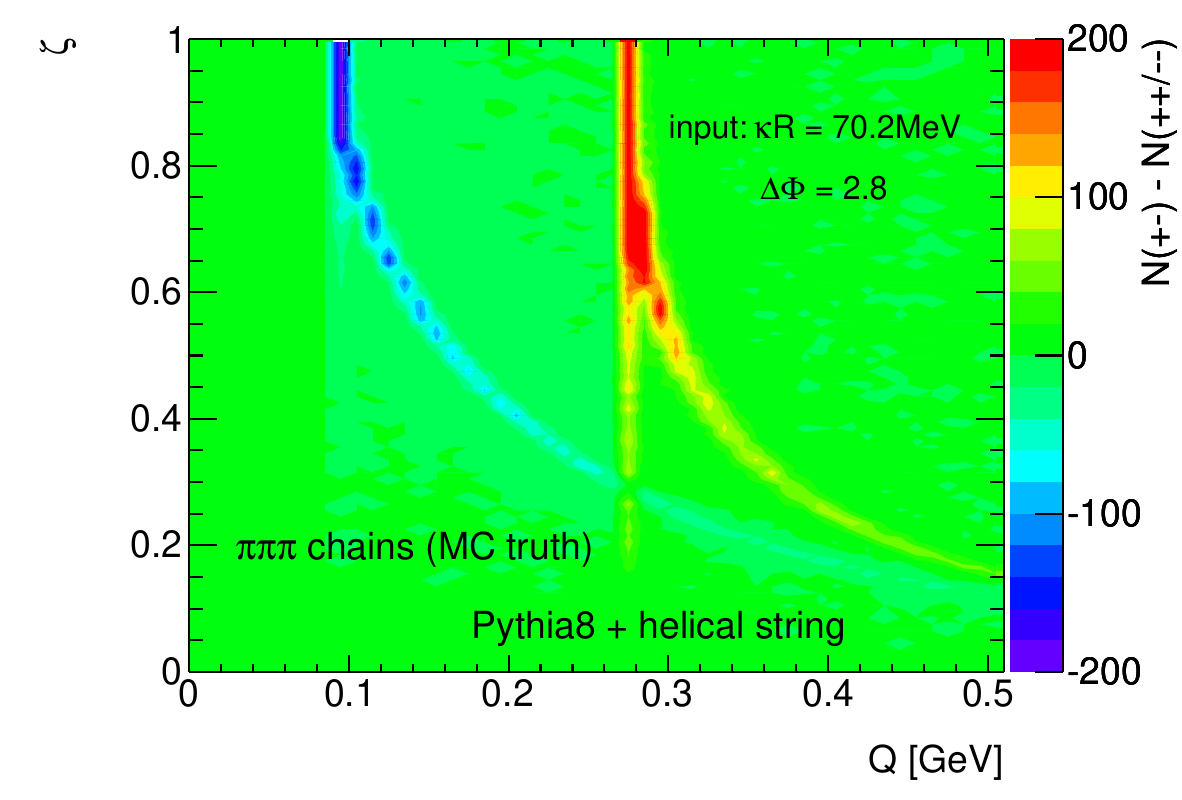}
  \includegraphics[width=0.45\textwidth]{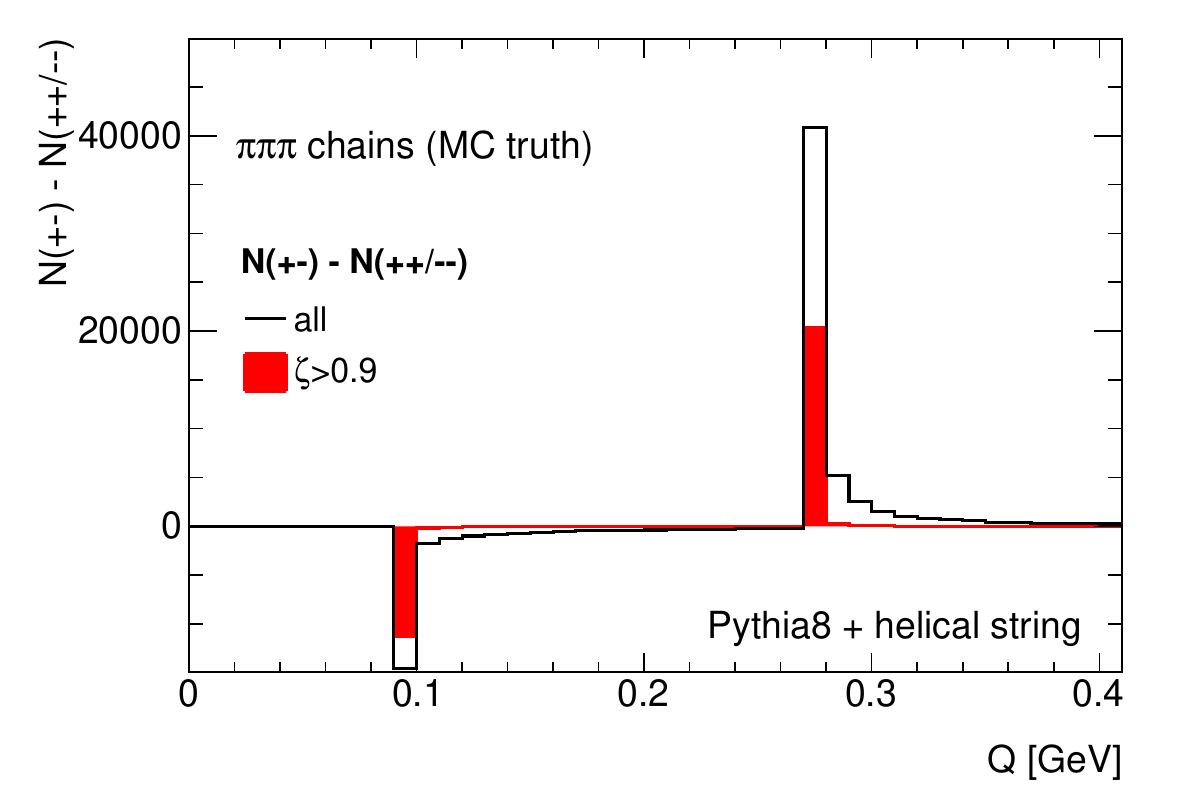}
  \caption{ Top: The $\zeta$ dependence of the momentum difference between pairs of direct pions drawn from a true triplet pion chain
   illustrates the effect of variable string pitch which results in non-negligible longitudinal momentum difference between pions.
   It is also obvious the smearing can be eliminated by measuring the momentum difference in the limit of $\zeta\rightarrow$ 1.
   Bottom:  Effect of $\zeta$ limit on the suppression of the tail in the measurement of properties of helical string with ideal transverse shape.
   The threshold value corresponds to the quantized transverse momentum difference, for $\kappa R$=70.2MeV and $\Delta\Phi$=2.8.
   \label{fig:hel_nosmear_zeta}}
   \end{center}
   \end{figure}

     Fig~\ref{fig:hel_nosmear_zeta} (top) illustrates the "running" of momentum difference as function of $\zeta$ for like-sign and opposite-sign
    pairs of pions ( pairs with rank difference 2 and 1, respectively, entered with positive and negative weights).
    It is obvious that in order to measure the threshold value
    of momentum difference (with vanishing longitudinal component), it is convenient to work in the limit of $\zeta\rightarrow 1$.
    The bottom plot shows the impact of selecting pairs with $\zeta>$0.9 on the suppression of the tail of both peaks (or, rather,
    dip and peak).  Both plots are done using ideal transverse helix shape with $\kappa R$=70.2 MeV and $\Delta\Phi$=2.8 ( please note
    the input values are chosen in a way which makes them consistent with mass of charged pion in the quantized fragmentation
    -- $m_{\pi} = \kappa R \sqrt{ \Delta\Phi^2 - (2\sin{\Delta\Phi/2})^2}$ -- an inconsistent input would results in biased output, too.)

  \begin{figure} [b]
  \begin{center}
  \includegraphics[width=0.45\textwidth]{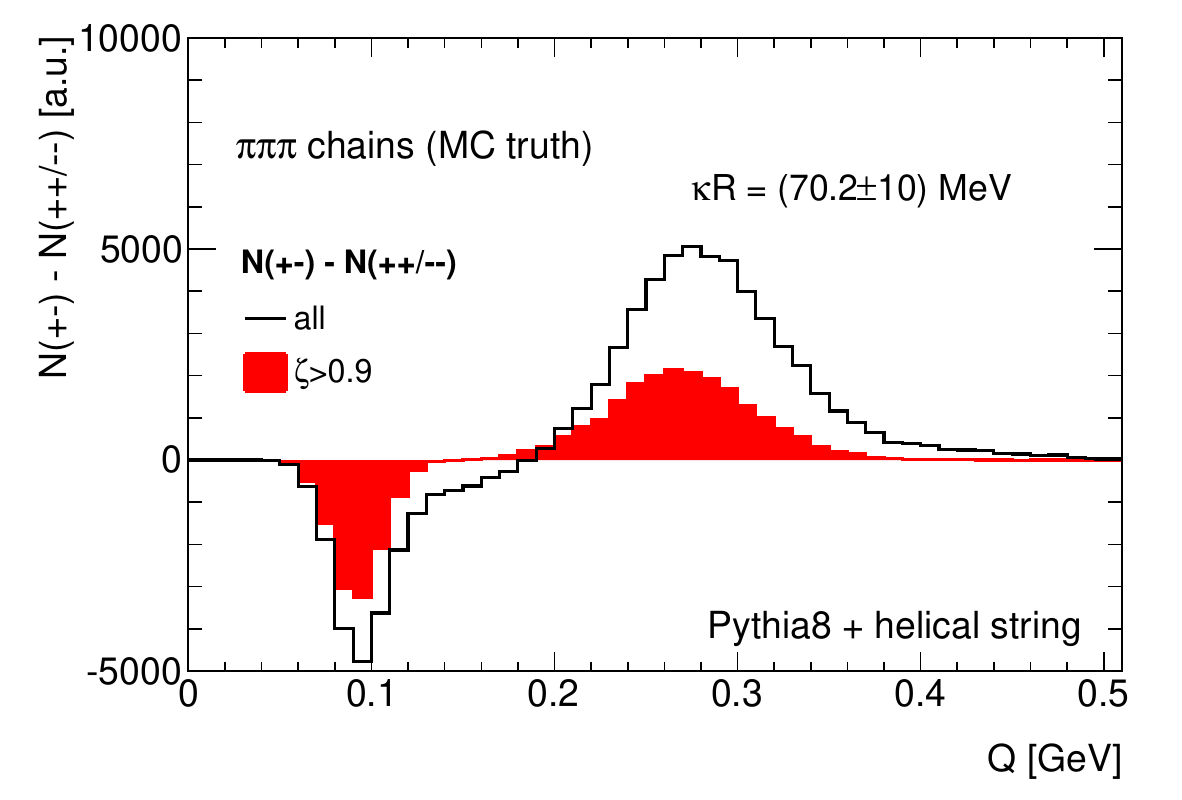}
  \includegraphics[width=0.45\textwidth]{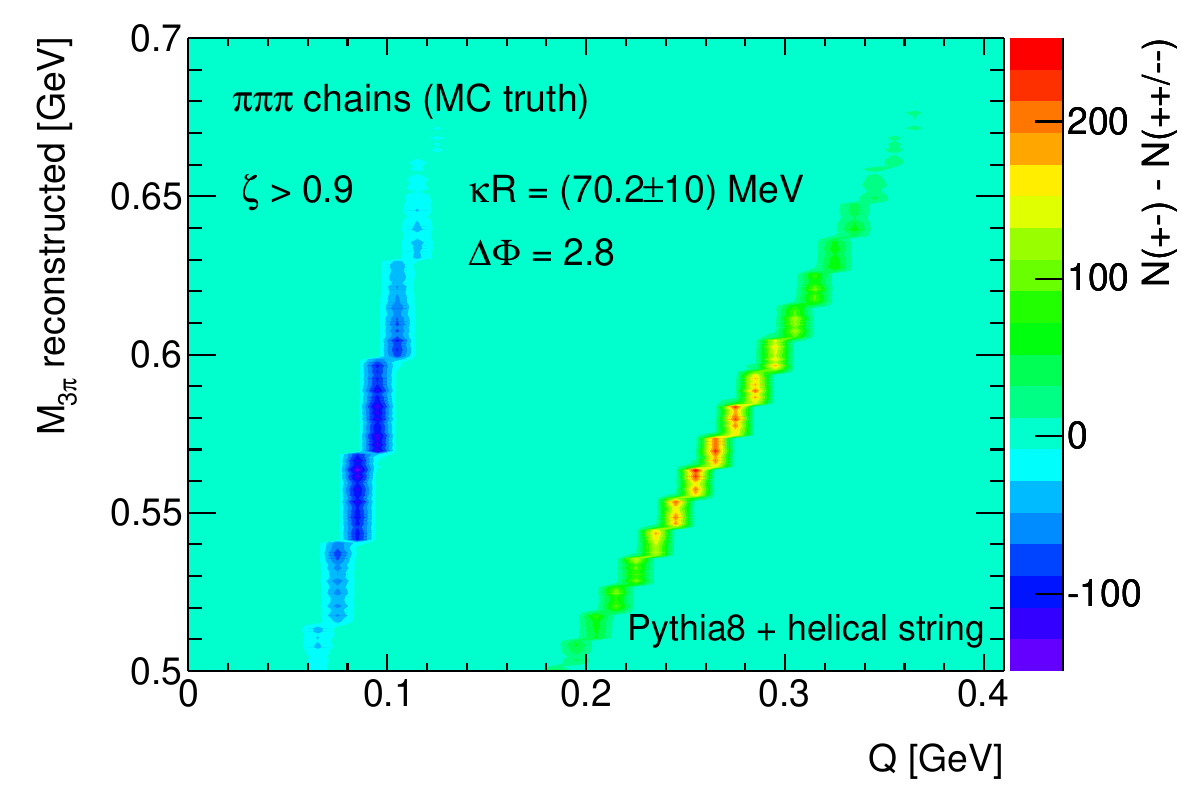}
  \caption{  Top:  Effect of $\zeta$ limit on the suppression of the tail in the measurement of properties of helical string with smeared transverse shape
   ( $\kappa R$=70.2$\pm$10MeV and $\Delta\Phi$=2.8). Bottom:  effect of smearing on the reconstructed pair momentum difference and reconstructed mass of triplet.
   \label{fig:hel_smear}}
   \end{center}
   \end{figure}

\begin{figure} [tbh!]
  \begin{center}
  \includegraphics[width=0.38\textwidth]{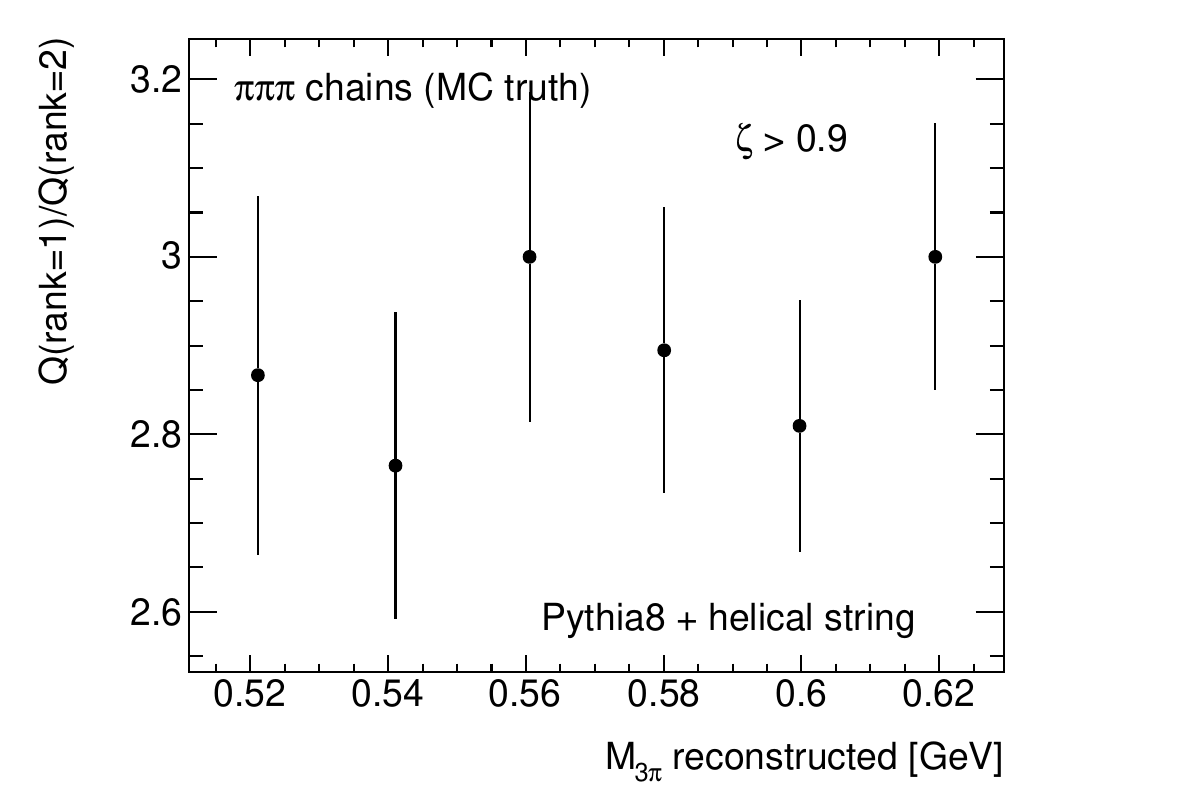}
  \includegraphics[width=0.38\textwidth]{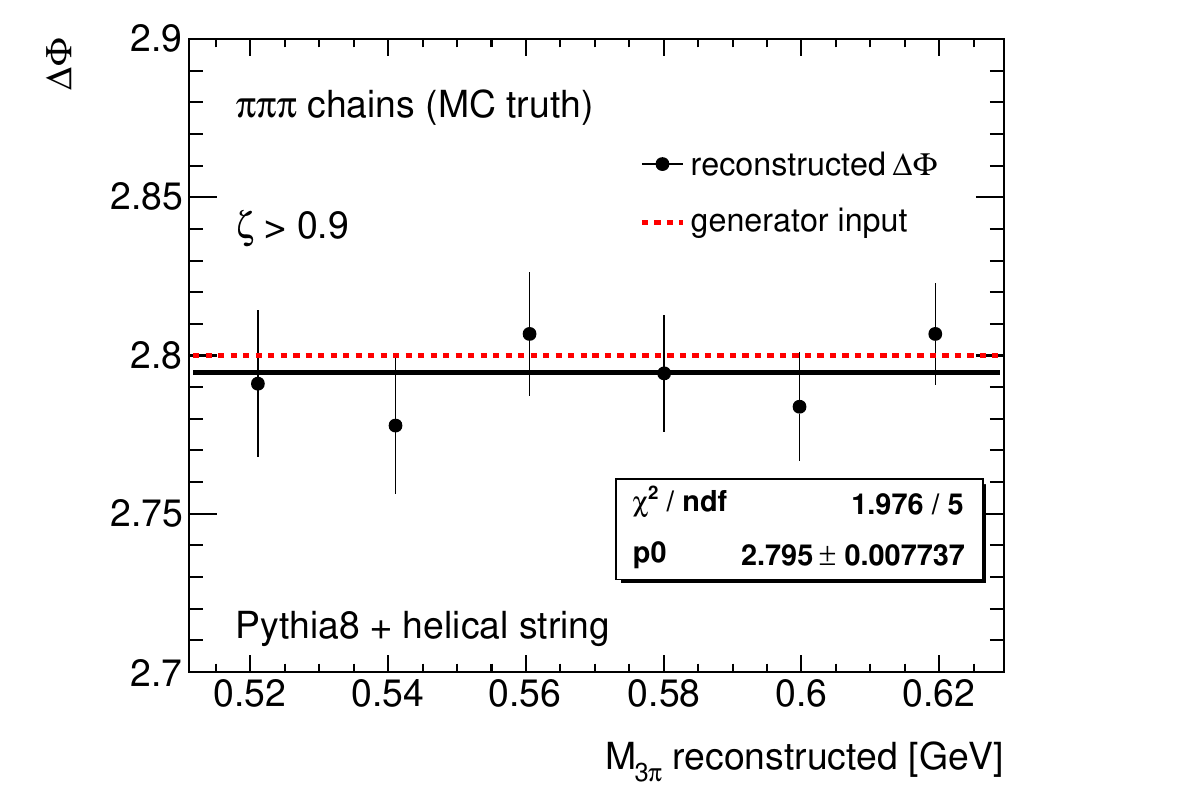}
  \includegraphics[width=0.38\textwidth]{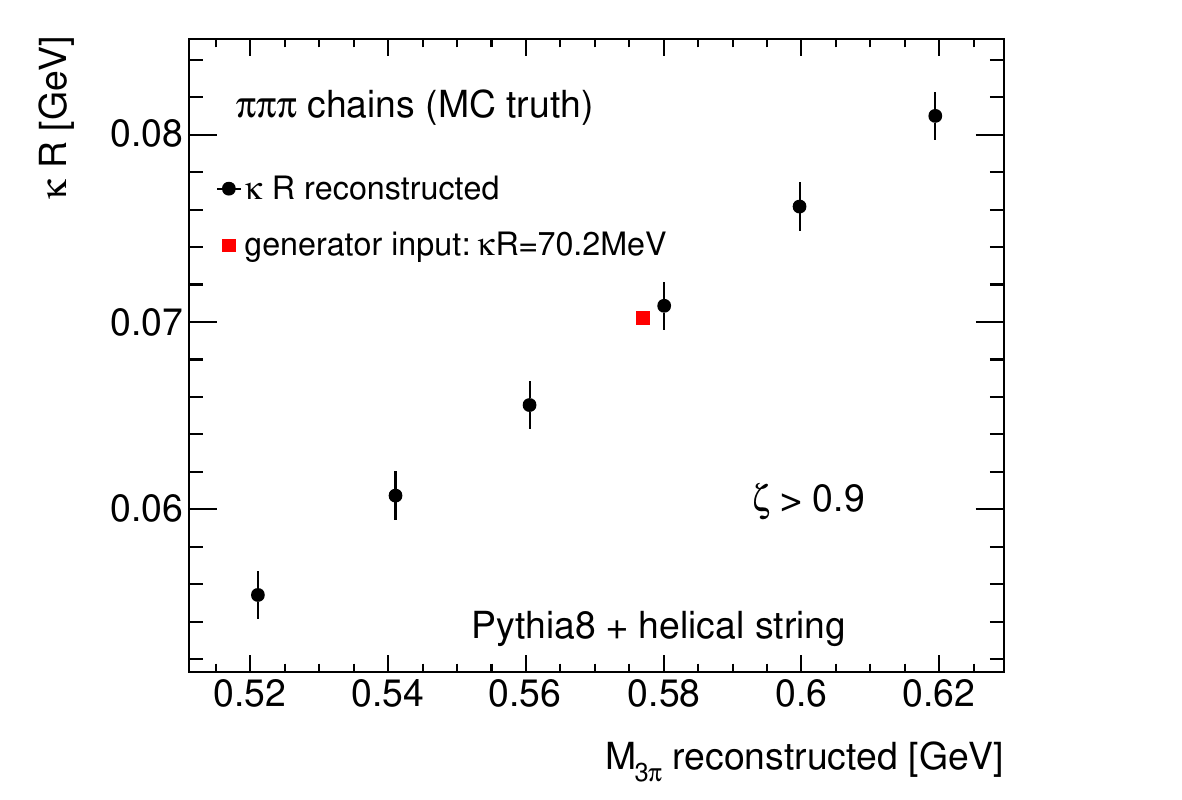}
   \includegraphics[width=0.4\textwidth]{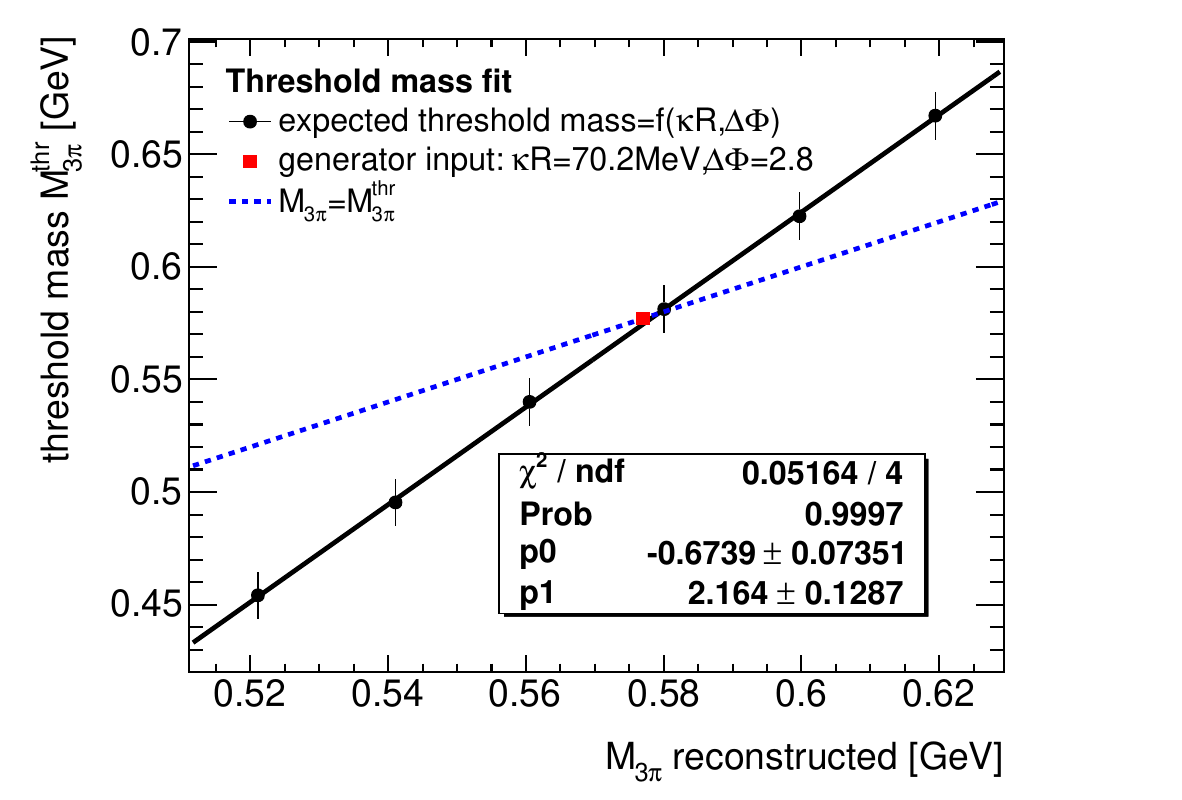}
 \caption{  From top to bottom:  Ratio $Q(\mathrm{r}=1)/Q(\mathrm{r}=2)$= sin($\Delta\Phi$/2)/sin($\Delta\Phi$) yields the measurement of $\Delta\Phi$,
  in good agreement with constant input value. Smearing of input $\kappa R$ is strongly correlated with the reconstructed triplet mass. The breakthrough
  sensitivity of the method resides in the possibility to check the self-consistency of results by comparing the expected threshold mass ( calculated
  from reconstructed $\kappa R, \Delta\Phi$ parameters)  with the reconstructed mass. The true threshold mass  lies where the expected
  mass coincides with the measured one, as shown on the bottom plot.  
   \label{fig:threshold_method}}
   \end{center}
   \end{figure}

     Smearing of the ideal helix shape, as well as various detector resolution effects, makes it difficult to decide where the true threshold lies.
    An illustration is provided in Fig.~\ref{fig:hel_smear} where reconstruction is done for high $\zeta$ values ($>$0.9). The smearing has been introduced
    by variation of $\kappa R$ according to a gaussian distribution with $\sigma$=10 MeV ( an example which is qualitatively conform
    to what is seen in real data, \cite{chains2}). The resulting mass spectrum of pion triplet chain
    is wide and it would be difficult to aspire on a precise measurement if not for the check of the internal consistency of the result which
    the over-constrained model offers.  The independent measurement of $Q(\mathrm{rank}=1,2) \rightarrow \kappa R, \Delta\Phi$
     can be put in concurrence with direct measurement
     of chain mass ( the threshold mass of triplet pion chain is  $m_{3\pi} = \kappa R \sqrt{ (3\Delta\Phi)^2 - (2\sin{3\Delta\Phi/2})^2}$ in the model of quantized fragmentation).  The sequence is shown in Fig.~\ref{fig:threshold_method}. The "measurement" is done by intervals of 20 MeV in the reconstructed triplet chain mass. The fit of the most likely value of $Q(\mathrm{rank}=1,2)$ is replaced by simple search of bin with maximal (minimal) content (distributions are narrow in this particular toy example), i.e. the value is put at the center of bin, with half of the bin size for the error. Ratio of rank 1 to rank 2 momentum difference provides the measurement of quantized helix phase $\Delta\Phi$ : $Q(\mathrm{r}=1)/Q(\mathrm{r}=2)$= sin($\Delta\Phi$/2)/sin($\Delta\Phi$).  No mass dependence is expected as this input parameter is kept fixed, and none is observed either. Fit with constant term is in good agreement with the input value. Variation of input $\kappa R$ parameter propagates to the variation of the reconstructed triplet mass.  The advantage of having an over constrained ( and predictive ) model
  consists in the possibility to apply a self-consistency check by requiring the measured triplet mass to coincide with the
   expected triplet threshold mass, calculated from measured $\kappa R$ and $\Delta\Phi$.  In this way, the precision of the measurement
   can be improved substantially.

\section{Long range correlations}

\begin{figure}[bh!]
\begin{center}
\includegraphics[width=0.45\textwidth]{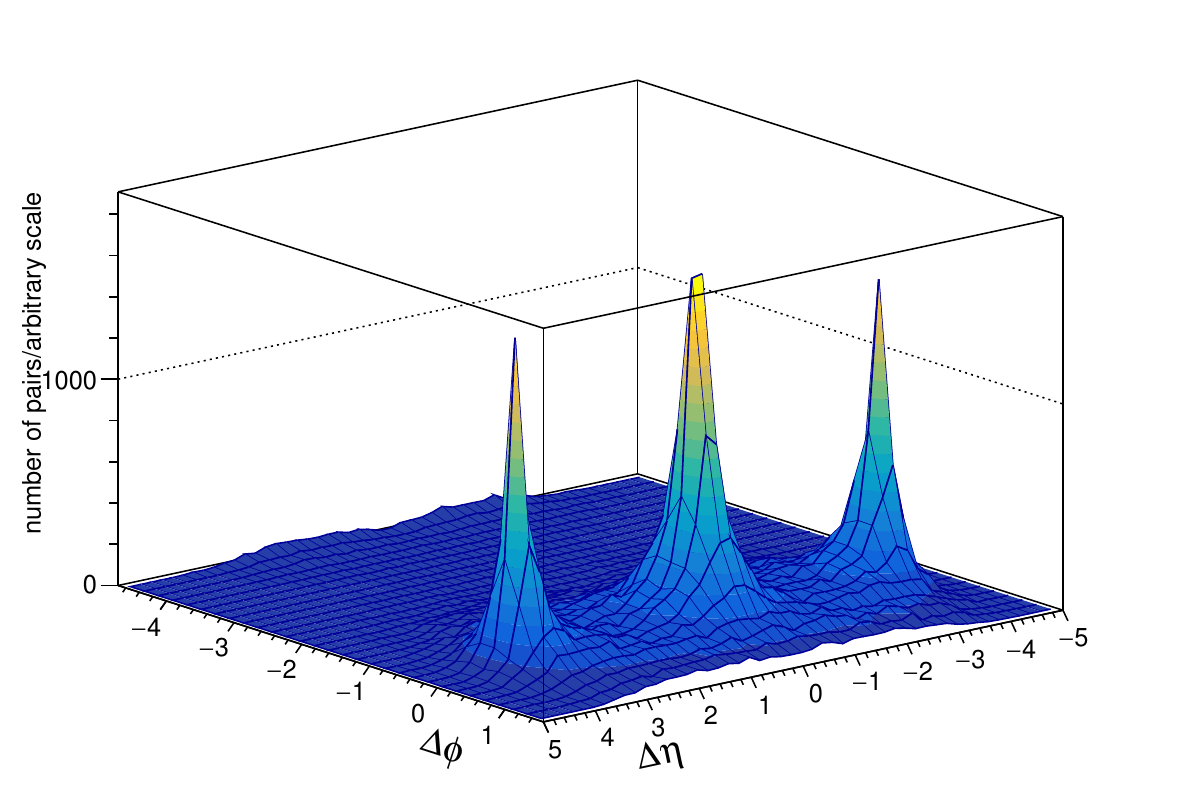}
\includegraphics[width=0.45\textwidth]{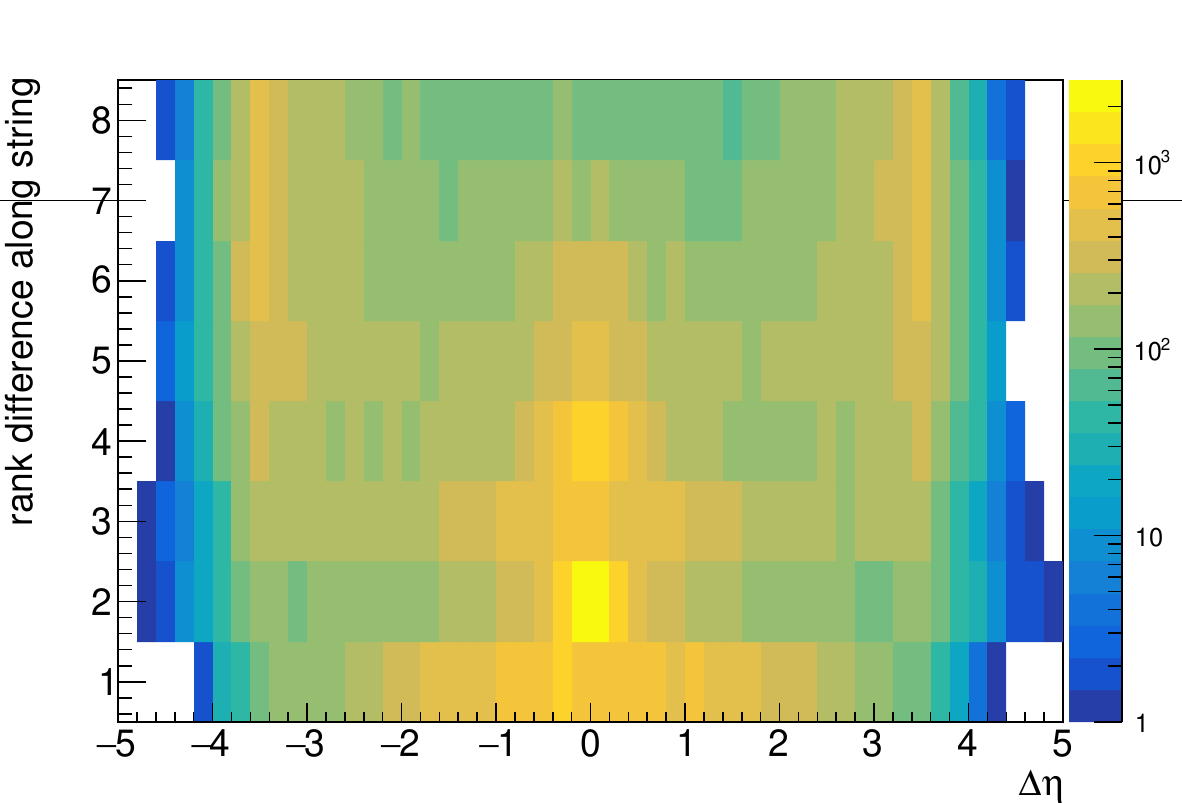}
\caption{ 
Top: The emergence of a ridge-like structure in the fragmentation of a helical string with a lateral boost. The narrow helix radius (0.07 fm)
implies a small intrinsic momentum ($\sim$140 MeV) of hadrons  and thus smaller smearing in the boosted direction. The recoiling system is not shown.
Bottom: The rank dependence of the ridge structure shown above. Higher eta difference is effectively dominated by pairs with larger
 rank distance (the distribution is truncated)  but the adjacent hadrons may have a large pseudorapidity difference as well. 
\label{fig:ridge}}
\end{center}
\end{figure}

  The presence of narrow strings may have been one of factors facilitating the emergence of the so-called ``ridge''
 effect observed both in HI and pp collisions, where azimuthal correlations spanning over a larger intervals of pseudorapidity
 appear in the data.
  
   The ridge-like structure can be easily generated by a laterally boosted helical string with parameters described
 above : the pions produced in the string fragmentation have instrinsic transverse momentum of only $\sim$ 140 MeV,
 w.r.t. the string axis, and are therefore likely to be measured in the direction of the boost. As an example,
 a simple string with the mass of 20 GeV has been generated, fragmented into set of pions within the helix string scenario,
 and boosted laterally (with $\gamma=$1.2). Fig.~\ref{fig:ridge} shows the angular correlations of the produced pions,
 with significant signal at large pseudorapidity difference. It is interesting to look at the rank dependence of these
 correlations (the bottom plot): the side peaks are dominated by larger rank differences, but adjacent pairs can be
 nevertheless observed at large pseudorapidity difference, extending up to 4.

\section{Further experimental input} 

    The model of the helical QCD string is rather successful in terms of the reduction of the number of independent parameters
  entering the simulation. For the validation and further development, it can be compared with a large variety
  of measurements (single particle spectra, particle correlations). The data used so far do not contain all the information
  which is necessary: the correlation measurements shown here do not identify the charged particles, and therefore
  do not see a difference between pions and kaons, for example. The properties of an ordered chain of kaons
  cannot be predicted easily within the model : kaons are expected to contain a string loop, and the properties of the
  string knotting are not known. The measurement of the $\Delta(Q)$
  for pairs of identified hadrons (pions,kaons,..)  is therefore of utmost interest.
  The model predicts the bulk of correlations should be carried by particles with (intrinsic) transverse momentum
  of $\sim$140 MeV, which is a value very close to, or even below, the acceptance of the LHC experiments.
  A dedicated measurement of the very low p$_T$  region would be very welcome indeed. 

\section{Conclusions}             
   The simulation strategy for the quantum properties of the string fragmentation is outlined and the first
  simulation of the so-called Bose-Einstein effect is performed, successfully reproducing the data. The correlation
  of like-sign pairs of hadrons is a consequence of non-trivial properties of the production of adjacent hadrons,
  which are themselves correlated, too.
  The building blocks of the simulation contributing to the description of the Q spectra are identified
  and the uncertainties discussed.
  The simulation marks an additional success by obtaining a signal of the azimuthal ordering of hadrons, even though
  the measurement is not reproduced in sufficient detail yet.
    From the initial studies it seems the long range correlations can be modelled by the fragmentation of boosted
  helical strings. A presence of adjacent hadron pairs in the long
  range correlation signal is not excluded.

\bibliography{chains_mc}

\end{document}